\documentclass[english,prl,two column]{revtex4-1}
\usepackage[LGR,T1]{fontenc}
\usepackage[latin9]{inputenc}
\usepackage{color}
\usepackage{bm}
\usepackage{amsmath}
\usepackage{amssymb}
\usepackage{graphicx}
\usepackage{esint}

\makeatletter

\DeclareRobustCommand{\greektext}{%
  \fontencoding{LGR}\selectfont\def\encodingdefault{LGR}}
\DeclareRobustCommand{\textgreek}[1]{\leavevmode{\greektext #1}}
\DeclareFontEncoding{LGR}{}{}
\DeclareTextSymbol{\~}{LGR}{126}
\newcommand{\lyxmathsym}[1]{\ifmmode\begingroup\def\b@ld{bold}
  \text{\ifx\math@version\b@ld\bfseries\fi#1}\endgroup\else#1\fi}

 \@ifundefined{textcolor}{}
 {%
   \definecolor{BLACK}{gray}{0}
   \definecolor{WHITE}{gray}{1}
   \definecolor{RED}{rgb}{1,0,0}
   \definecolor{GREEN}{rgb}{0,1,0}
   \definecolor{BLUE}{rgb}{0,0,1}
   \definecolor{CYAN}{cmyk}{1,0,0,0}
   \definecolor{MAGENTA}{cmyk}{0,1,0,0}
   \definecolor{YELLOW}{cmyk}{0,0,1,0}
 }

\makeatother

\usepackage{babel}
\begin{document}

\title{Scalable quantum simulation of pulsed entanglement and Einstein-Podolsky-Rosen
steering in optomechanics}

\author{S. Kiesewetter$^{1}$, Q. Y. He$^{1}$, P. D. Drummond$^{1}$, and
M. D. Reid$^{1}$ }

\affiliation{$^{1}$Centre for Quantum Atom Optics, Swinburne University of Technology,
Melbourne, Australia\\
}





\begin{abstract}
We demonstrate a complete, probabilistic quantum dynamical simulation
of the standard nonlinear Hamiltonian of optomechanics, including
decoherence at finite temperatures. Robust entanglement of an optical
pulse with the oscillator is predicted, as well as strong quantum
steering between the optical and mechanical systems. Importantly,
our probabilistic quantum simulation method uses the positive-P technique,
which is scalable to large Hilbert spaces.
\end{abstract}
\maketitle
\textbf{Introduction.} Optomechanical oscillators provide a fundamental
test of mesoscopic quantum mechanics, as well as having potential
technological applications in a wide variety of sensitive measurements.
Impressive success in cooling optomechanical systems near their ground
state has been reported experimentally \cite{cooling_Connell_2010_Nature,cooling_Chan_2011_Nature,cooling_Gro_2009_NP,cooling_Teufel},
resulting in the demonstration of a number of quantum mechanical effects
for mesoscopic systems \cite{quantum_macro,quantum_macro2,quantum_macro3,quantum_macro4}.
An outstanding goal is to observe quantum correlations, such as entanglement
and the nonlocality predicted by Einstein, Podolsky and Rosen (EPR)
\cite{EPR1935,Reid1989} for macroscopic, massive objects \cite{Schrodinger,ent-opto-Giovannetti,ent-opto-Mancini,Simon}.
In the first instance, it is of interest whether an optical field
can entangle with a massive oscillator, and whether the two systems
can show the strange directional ``spooky action-at-a-distance''
\cite{Born} effects that Schrodinger called ``steering'' \cite{Schrodinger-steer,steering_wiseman,cava_steering2009}.

In this paper, we carry out the first scalable, probabilistic quantum
mechanical simulations of the standard nonlinear optomechanical Hamiltonian.
This helps to unravel the quantum mechanical interplay between entanglement
generation, created by the nonlinear radiation coupling, and the thermal
decoherence due to reservoirs. We use this to study the dynamical
generation of entanglement and EPR-steering correlations, for pulsed
inputs and realistic experimental parameters. Models used previously
often make numerous assumptions, ranging from linearization \cite{ent-opto-Giovannetti}
to adiabatic approximations \cite{ent-opto-Hofer}, or both. While
very useful in giving analytic results, it unclear how justified these
assumptions are. In our simulations we utilize the exact positive-P
phase-space method \cite{gen-p-rep}, which exists as a positive probability
distribution for all quantum states. 

The main limitation in current optomechanics experiments is that long
interaction times lead to increased decoherence, owing to a coupling
to the environmental heat bath which is often at relatively high temperature.
The use of pulsed probes can overcome this, and it was recently proposed\cite{ent-opto-Hofer}
to create and verify entanglement with two successive pulses of light
so that interactions can take place on fast time scales \cite{fast-coupling-Vanner2011,fast-coupling-Pikovski-np2012}.
This theoretical treatment was restricted to an adiabatic, linearized
study using symmetric entanglement measures \cite{DGCZ}. They showed
that entanglement is feasible, provided $Qf\gg k_{B}T_{bath}/h$,
where $T_{bath}$ is the temperature of the environment, $f$ is the
frequency of mechanical oscillator and $Q$ is the cavity quality.
By comparison, our method can treat arbitrary pulse shapes, is not
linearized, and uses optimized measures for entanglement and EPR.
We compare our results with the truncated Wigner method \cite{Wigner,Wigner_Castin,Wigner_Steel},
valid for large photon-numbers \cite{Peterepl1993,Peter2002,Dechoum2004}. 

This simulation technique, which has been used on other parametric
problems \cite{Reid-Drumm}, has no approximations apart from those
inherent in the standard model \cite{Pace-Collett-Walls}. It successfully
produces results for realistic experimental parameters at large photon
number. Compared to direct diagonalization \cite{Ludwig} or quantum
trajectory approaches \cite{Kronwald}, the method readily scales
to large Hilbert spaces \cite{Carter}. Neither approximations \cite{Qiu}
nor new hardware \cite{LLoyd} are required, apart from sampling error
issues with high-order correlations \cite{Opanchuk}. Therefore, it
has excellent potential for treat the challenging new generation of
multimode optomechanical crystal devices \cite{Chan-optocrystal}.

The results of this full analysis validate some of the earlier predictions
for the simplified model in Refs. \cite{optoHe&Reid,ent-opto-Hofer}.
It is directly useful for addressing experiments with optomechanical
systems. Our main finding is that entanglement of pulse and oscillator
mode proves to be robust, without requiring a low temperature reservoir,
but that the thermal noise provides a stronger barrier to the EPR
steering paradox than to entanglement. The thermal barrier acts directionally,
to prevent steering of the mechanical system when thermally excited.
This is a fundamentally asymmetric manifestation of nonlocality, beautifully
illustrated by the oscillator-pulse system. 

\begin{figure}
\begin{centering}
\includegraphics[width=0.8\columnwidth]{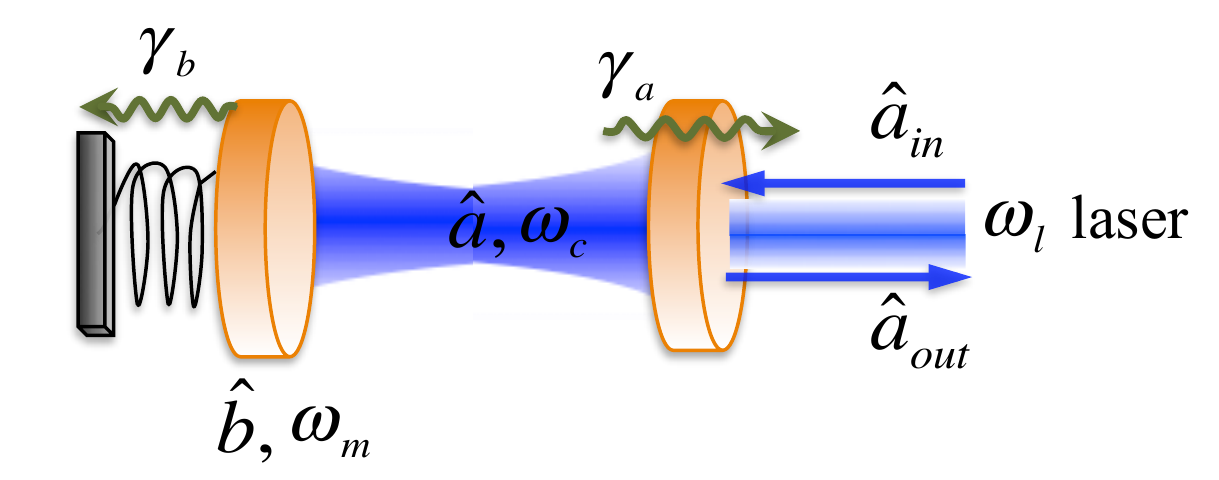}
\par\end{centering}

\caption{Schematic diagram of a driven optomechanical system. Here, $\hat{a}_{in}$
and $\hat{a}_{out}$ denote ingoing and outgoing fields, while $\hat{a}$
is an annihilation operator of the optical mode with resonance frequency
$\omega_{c}$, and $\hat{b}$ corresponds to a mechanical mode with
oscillation frequency $\omega_{m}$. A light pulse of duration $\tau$
and carrier frequency $\omega_{l}$ impinges on the cavity and interacts
with the mirror via radiation pressure. \label{fig:diagram}}
\end{figure}

\textbf{Hamiltonian and stochastic equations.} We consider the standard,
single-mode model \cite{Hamil_Brag,Hamil_pierre,Pace-Collett-Walls}
for an optomechanical Fabry-Perot cavity with coherent pumping and
damping. The Hamiltonian includes the energy of the cavity mode at
angular frequency $\omega_{c}$, the mechanical oscillator mode at
$\omega_{m}$, and an input driving the cavity mode with amplitude
$E(t)$: 
\begin{eqnarray}
\hat{H}/\hbar & = & \omega_{c}\hat{a}^{\dagger}\hat{a}+\omega_{m}\hat{b}^{\dagger}\hat{b}+\chi_{0}\hat{a}^{\dagger}\hat{a}(\hat{b}+\hat{b}^{\dagger})\nonumber \\
 &  & +iE(t)(\hat{a}^{\dagger}e^{-i\omega_{l}t}-\hat{a}e^{i\omega_{l}t})+\hat{H}_{r},\label{eq:Full_H}
\end{eqnarray}
The first two terms give the energy of the cavity field and the mechanical
oscillator, the third term denotes the optomechanical interaction,
where $\chi_{0}$ is the single-photon coupling, and the fourth is
the coupling to the coherent input field. $\hat{H}_{r}$ describes
dissipation of the two subsystems via coupling to reservoirs. The
system is driven by a light pulse of duration $\tau$ and carrier
frequency $\omega_{l}=\omega_{c}-\Delta$, with $N_{ph}$ photons.
The laser driving strength is $E(t)=E_{0}\varepsilon(t)=\sqrt{2\gamma_{a}N_{ph}}\varepsilon(t)$,
and the envelope function $\varepsilon(t)$ is normalized so that
$\int_{0}^{\tau}dt|\varepsilon(t)|^{2}=1$ . A diagram is shown in
Fig. \ref{fig:diagram}. 

This is a driven open system, hence the density matrix must be calculated
as the solution of a master equation. For simplicity, we transform
to a rotating frame in which the free-field time evolution of the
field operators is removed, where $\Delta=\omega_{c}-\omega_{l}$
is the detuning of the laser driving frequency with respect to the
cavity resonance. The master equation for the reduced density operator
is then given by
\begin{eqnarray}
\frac{d\hat{\rho}}{dt} & = & -i[\Delta\hat{a}^{\dagger}\hat{a}+\omega_{m}\hat{b}^{\dagger}\hat{b}+\chi_{0}\hat{a}^{\dagger}\hat{a}(\hat{b}+\hat{b}^{\dagger})+iE(t)(\hat{a}^{\dagger}-\hat{a}),\hat{\rho}]\nonumber \\
 &  & {+\sum_{i=1,2}\gamma_{i}(\bar{n}_{i,th}+1)(2\hat{a}_{i}\hat{\rho}\hat{a}_{i}^{\dagger}-\hat{a}_{i}^{\dagger}\hat{a}_{i}\hat{\rho}-\hat{\rho}\hat{a}_{i}^{\dagger}\hat{a}_{i})}\nonumber \\
 &  & {+\sum_{i=1,2}\gamma_{i}\bar{n}_{i,th}(2\hat{a}_{i}^{\dagger}\hat{\rho}\hat{a}_{i}-\hat{a}_{i}\hat{a}_{i}^{\dagger}\hat{\rho}-\hat{\rho}\hat{a}_{i}\hat{a}_{i}^{\dagger})\,\,.}\label{eq:master}
\end{eqnarray}
where $\gamma_{a}$ and $\gamma_{b}$ are the cavity decay rate and
mechanical dissipation rate. Here the vector $\hat{\bm{a}}=(\hat{a}_{1},\hat{a}_{2})=(\hat{a},\hat{b})$
is introduced for convenience, with $i=1,2\sim a,b$ indexing the
optical and mechanical modes respectively. 

We transcribe this master equation into a $c-$number phase space
evolution equation, with similar features to related equations found
in optical fibre simulations \cite{Peterepl1993}. First we use the
positive-P representation \cite{gen-p-rep}, which gives an exact,
positive phase-space representation of any quantum state (\ref{eq:master}).
This method has a normally-ordered quantum-stochastic correspondence,
with $\langle\alpha^{+}\alpha\rangle_{S}=\langle\hat{a}^{\dagger}\hat{a}\rangle_{Q}$
, $\langle\beta^{+}\beta\rangle_{S}=\langle\hat{b}^{\dagger}\hat{b}\rangle_{Q}$.
After obtaining a Fokker-Planck equation, one obtains a set of four
complex $It\hat{o}$ stochastic equations:
\begin{eqnarray}
d\alpha & = & \left\{ E(t)-\left[i\Delta+i\chi_{0}(\beta+\beta^{+})+\gamma_{a}\right]\alpha\right\} dt+dW_{a},\nonumber \\
d\beta & = & \left[-\left(i\omega_{m}+\gamma_{b}\right)\beta-i\chi_{0}\alpha\alpha^{+}\right]dt+dW_{b},\nonumber \\
d\alpha^{+} & = & \left\{ E^{*}(t)+\left[i\Delta+i\chi_{0}(\beta+\beta^{+})-\gamma_{a}\right]\alpha^{+}\right\} dt+dW_{a}^{+},\nonumber \\
d\beta^{+} & = & \left[\left(i\omega_{m}-\gamma_{b}\right)\beta^{+}+i\chi_{0}\alpha\alpha^{+}\right]dt+dW_{b}^{+},\label{eq:positive-P}
\end{eqnarray}
The noises here are due to both internal nonlinearities and thermal
noise inputs, so that generically $dW_{i}=dW_{i}^{\chi}+\sqrt{2\gamma_{i}}dW_{i}^{in}$,
whose non-vanishing correlations are: $\langle dW_{i}^{in}dW_{i}^{in+}\rangle=\bar{n}_{i,th}dt$,
$\langle dW_{i}^{\chi}W_{j}^{\chi}\rangle=-i\delta_{i,3-j}\chi_{0}\alpha dt$,
and $\langle dW_{i}^{\chi+}W_{j}^{\chi+}\rangle=i\delta_{i,3-j}\chi_{0}\alpha^{+}dt$.\textcolor{red}{{} }

As an alternative approach which is simpler -- but approximate --
we can use the truncated Wigner distribution \cite{Wigner,Wigner_Castin,Wigner_Steel},
which is a symmetrically ordered representation. After removing higher-order
derivatives in the Fokker-Planck equation (in an approximation valid
at large photon number), we obtain the equations:
\begin{eqnarray}
d\alpha & = & \left\{ E(t)-\left[i\Delta+i\chi_{0}(\beta+\beta^{*})+\gamma_{a}\right]\alpha\right\} dt+\sqrt{2\gamma_{a}}dW_{a}^{in},\nonumber \\
d\beta & = & \left[-\left(i\omega_{m}+\gamma_{b}\right)\beta-i\chi_{0}|\alpha|^{2}\right]dt+\sqrt{2\gamma_{b}}dW_{b}^{in}.\label{eq:wigner}
\end{eqnarray}
Here there are two complex variables, with non vanishing noise correlations
are given by $\langle dW_{i}^{in}\rangle=0,$ $\langle dW_{i}^{in}dW_{i}^{in*}\rangle=(\bar{n}_{i,th}+1/2)dt$,
where $i=a,b$, and $\bar{n}_{i,th}$ are the mean heat bath occupations.
These equations imply that $\langle\alpha\alpha^{\dagger}\rangle_{S}=\langle\hat{n}+1/2\rangle_{Q}=1/2$
when there is no driving or coupling, as required for a vacuum state. 

In the simulations, we drive the cavity with a blue-detuned laser
pulse where the resonant scattering to the lower (Stokes) sideband
($\omega_{c}=\omega_{l}-\omega_{m}$, i.e. $\Delta=-\omega_{m}$)
enhances entanglement. We choose parameters that correspond to recent
experiments on $Si$ optomechanical crystal structures \cite{cooling_Chan_2011_Nature},
with $\omega_{m}/2\pi=3.7GHz$, $Q_{m}=\omega_{m}/\gamma_{b}=10^{5}$,
$\gamma_{b}/2\pi=37KHz$, $\gamma_{a}/2\pi=0.26GHz$, $\chi_{0}/2\pi=910KHz$.
The pulse parameters used a total photon number of $N_{ph}=8.2\times10^{6}$
and a pulse duration of ${\tau=0.04\mu s}$. We present results for
square pulses here, although a variety of pulse shapes $\varepsilon(t)$,
ranging from square waves to Gaussians, all gave strong entanglement
and steering. We will describe these more general results elsewhere,
for space reasons. 

Two different heat bath temperatures were chosen for comparison purposes:
either with a cold reservoir at $T_{bath}=200mK$, or a hot reservoir
of $T_{bath}=20K$ as in recent experiments. In both cases, the initial
mechanical occupation number was chosen as $n_{b,0}=0.7$, corresponding
to an initial oscillator temperature of $200mK$. This is typically
obtained using laser pre-cooling of the mechanical oscillator. 

To study the optical properties of the output field, it is convenient
to use operators in a frame rotating with $\lyxmathsym{\textgreek{w}}_{m}$,
by utilizing $\hat{a}^{r}=\hat{a}e^{-i\omega_{m}t}$ and $\hat{b}^{r}=\hat{b}e^{i\omega_{m}t}$,
with similar definitions for the input and output operators. A normalized
output mode in the rotating frame is obtained  using the standard
cavity input-output relations, $\hat{a}_{out}(t)+\hat{a}_{in}(t)=\sqrt{2\gamma_{a}}\hat{a}^{r}$
\cite{inout_Yurke,inout-Gardiner-Collett}, together with a mode function
designed to match the gain characteristics of the cavity \cite{ent-opto-Hofer}.
We define 
\begin{eqnarray}
\hat{A}_{out} & = & \sqrt{\frac{1}{\mathcal{N}\left(\tau\right)}}\int_{0}^{\tau}dte^{r(t)}\hat{a}_{out}(t),
\end{eqnarray}
in terms of the integrated gain, $r(t)=\int_{0}^{t}G(t')dt'$ and
normalization, $\mathcal{N}\left(\tau\right)=\int_{0}^{\tau}e^{2r(t)}dt$,
where $G(t)=\sqrt{\chi_{0}^{2}E(t)/\gamma_{a}(\Delta^{2}+\gamma_{a}^{2})}$
is the effective optomechanical coupling. This causes the mirror motion
to become correlated with a light mode of central frequency $\omega_{c}=\omega_{l}-\omega_{m}$. 

In either representation, the corresponding stochastic equation for
the output field is 
\begin{eqnarray}
{d\alpha_{out}} & {=} & {\sqrt{\frac{1}{\mathcal{N}\left(\tau\right)}}e^{r(t)}\left(\sqrt{2\gamma_{a}}\alpha^{r}dt-dW_{a}^{in,r}\right).}
\end{eqnarray}

\textbf{Robust asymmetric entanglement }- The measured output covariance
of a general quadrature $\hat{X}_{c}^{\theta,out}=(e^{-i\theta}\hat{A}_{out}(\tau)+e^{i\theta}\hat{A}_{out}^{\dagger}(\tau))/2$,
$\hat{X}_{m}^{\varphi,out}=(e^{-i\varphi}\hat{b}(\tau)+e^{i\varphi}\hat{b}^{\dagger}(\tau))/2$
can be written as 
\begin{eqnarray}
V_{cm}^{\theta,\varphi} & = & \langle\Delta\hat{X}_{c}^{\theta}\Delta\hat{X}_{m}^{\varphi}\rangle.
\end{eqnarray}
Here the fluctuations $\Delta\hat{X}_{c,m}^{\theta,\varphi}$ are
defined as $ $$\Delta\hat{X}_{c,m}^{\theta,\varphi}=\hat{X}_{c,m}^{\theta,\varphi}-\langle\hat{X}_{c,m}^{\theta,\varphi}\rangle$,
where $\theta,\varphi$ are the phase angles for a phase-sensitive
local oscillator measurement. We then calculate continuous variable
entanglement and EPR correlations between the output optical field
quadrature and the position and momentum of the resonator. The optimal
phase to obtain the strongest signature of entanglement is ${\theta,\varphi=0,\pi/2}$. 

\begin{figure}
\begin{centering}
\includegraphics[width=0.9\columnwidth]{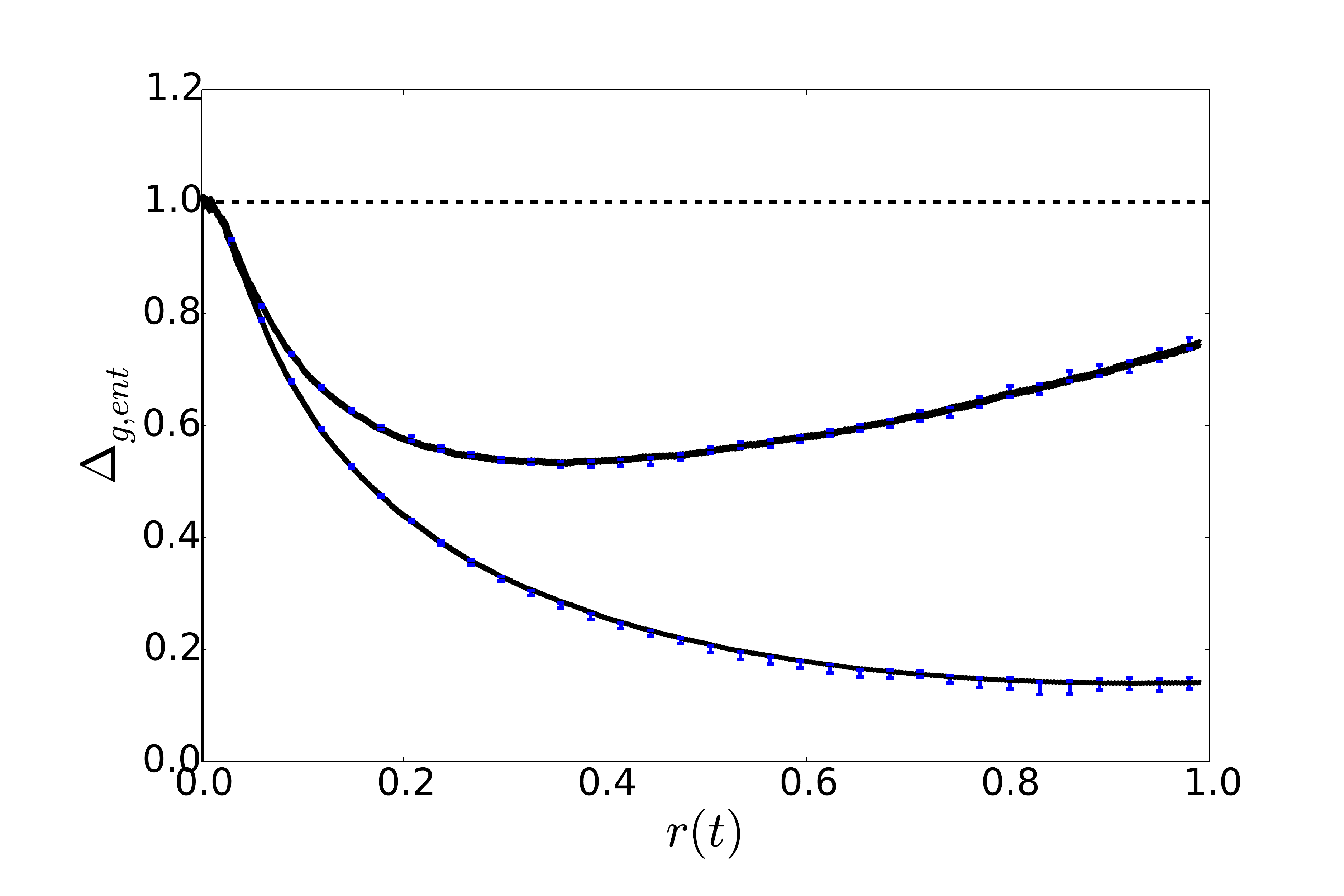}
\par\end{centering}

\caption{The entanglement signature $\Delta_{g,ent}$ is plotted versus $r(t)$,
for an initial mechanical occupation number $n_{b,0}=0.7$ and two
values of heat bath occupation $\bar{n}_{b,th}=0.7(lower),\ 112(upper)$
for the oscillator. The cavity bath is assumed a vacuum, i.e. $n_{a,0}=\bar{n}_{a,th}=0$.
Here the laser driving uses a square pulse shape ($\varepsilon(t)=1/\sqrt{\tau}$).
Solid lines represent the truncated Wigner results with $80000$ trajectories.
Error bars represent sampling errors using the same trajectory numbers
but with the exact positive P-representation method. \label{fig:asy-ent}}
\end{figure}

We first consider the entanglement signatures. This is best indicated
using asymmetric weightings of the quadratures ($\hbar=1/2$) \cite{optoHe&Reid}
\begin{equation}
\Delta_{g,ent}=\frac{\left[\Delta(\hat{X}_{m}^{out}+g_{x}\hat{X}_{c}^{out})\right]^{2}+\left[\Delta(\hat{P}_{m}^{out}+g_{p}\hat{P}_{c}^{out})\right]^{2}}{\left(1+g_{x}g_{p}\right)/2}<1,\label{eq:asy_ent}
\end{equation}
where $g_{x},g_{p}$ are real gains used in post-processing the data.
In this system, the $\Delta_{g,ent}$ can be minimized by choosing
the optimal gain factor $g_{x,p}=g$
\begin{equation}
g=\frac{-b+\sqrt{b^{2}-4ac}}{2a},\label{eq:optimalg}
\end{equation}
where $c=\langle\hat{X}_{m}^{out},\hat{X}_{c}^{out}\rangle+\langle\hat{X}_{c}^{out},\hat{X}_{m}^{out}\rangle=-a$
($\langle\hat{k},\hat{l}\rangle=\langle\hat{k}\hat{l}\rangle-\langle\hat{k}\rangle\langle\hat{l}\rangle$)
and $b=2[(\Delta\hat{X}_{c}^{out})^{2}-(\Delta\hat{X}_{m}^{out})^{2}]$.
For the case of a symmetric distribution in both quadratures, $\Delta\hat{X}_{m}^{out}=\Delta\hat{P}_{m}^{out},\ \Delta\hat{P}_{c}^{out}=\Delta\hat{X}_{c}^{out}$,
one obtains $g_{x}=-g_{p}$. We assume the initial state of the light
field to be the vacuum state, $\Delta\hat{X}_{c}^{in}=\Delta\hat{P}_{c}^{in}=1/2$,
and that of the mirror to be a thermal state with mean excitation
number $n_{b,0}$, so that $\Delta\hat{X}_{m}^{in}=\Delta\hat{P}_{m}^{in}=\sqrt{n_{b,0}/2+1/4}$.
The resulting predictions for entanglement detected by the asymmetric
witnesses (\ref{eq:asy_ent}) are plotted in Fig. \ref{fig:asy-ent}.

Using these entanglement signatures, we have simulated the robust
asymmetric EPR paradox recently presented in Ref. \cite{optoHe&Reid},
but for the full standard optomechanical model, without any assumptions.
A graph of the predicted the quantum entanglement with low and high
temperature is given in Fig. \ref{fig:asy-ent}. These calculations
completely simulate recent experiments on $Si$ optomechanical crystal
structures \cite{cooling_Chan_2011_Nature}. They also demonstrate
excellent agreement between the exact positive P-representation and
the approximate Wigner method for these parameter values. 

The physical interpretation of these results is that for any given
initial mechanical oscillator occupation number $n_{b,0}$, we can
always obtain entanglement for $r=\int_{0}^{t}G(t')dt'>0$, provided
one uses the asymmetric criteria (\ref{eq:asy_ent}) and selects an
optimal choice of gain factor $g$ (\ref{eq:optimalg}). This means
we can detect entanglement in the presence of thermal mechanical decoherence,
without the need to use laser cooling to reduce the value of $n_{b,0}$.
However, the minimum value $r_{0}$ required for entanglement detection
depends on the accuracy achieved in selecting the gain factors $g$. 

Figure \ref{fig:asy-ent} indicates entanglement at a temperature
$T_{bath}=20K$ ($\bar{n}_{b,th}=112$), provided the oscillator is
pre-cooled to $T_{0}=200mK$ ($n_{b,0}\sim0.7$). This is sensitive
to the occupation number $\bar{n}_{b,th}$ of the mechanical heat
bath, but is more robust to thermal effects than using the symmetric
criterion in Ref. \cite{ent-opto-Hofer}. 

\textbf{EPR correlations and one-way steering.} To signify the EPR
steering paradox, it is not sufficient to simply demonstrate entanglement.
Here, we present realistic predictions for the steering, optimized
for $g$, using phase-space quantum simulations. A simple way to determine
if an EPR steering paradox is realized is by testing whether a Heisenberg
uncertainty relation is \textquotedblleft{}violated\textquotedblright{}
by inferring uncertainties under the assumptions of local realism
(LR), so that \cite{rmp_reid,Reid1989}
\begin{align}
E_{m|c} & =4\Delta_{inf}\hat{X}_{m}^{out}\Delta_{inf}\hat{P}_{m}^{out}\nonumber \\
 & =4\Delta(\hat{X}_{m}^{out}+g_{x}\hat{X}_{c}^{out})\Delta(\hat{P}_{m}^{out}+g_{P}\hat{P}_{c}^{out})<1.\label{eq:steer_m|c}
\end{align}
Here $\Delta_{inf}\hat{X}_{m}^{out}$, $\Delta_{inf}\hat{P}_{m}^{out}$
are the inferred uncertainties, and the optimal scale factors are
given as $g{}_{x}=-(\langle\hat{X}_{m}^{out},\hat{X}_{c}^{out}\rangle+\langle\hat{X}_{c}^{out},\hat{X}_{m}^{out}\rangle)/2(\Delta\hat{X}_{c}^{out})^{2}$
and $g{}_{p}=-(\langle\hat{P}_{m}^{out},\hat{P}_{c}^{out}\rangle+\langle\hat{P}_{c}^{out},\hat{P}_{m}^{out}\rangle)/2(\Delta\hat{P}_{c}^{out})^{2}$.
We note that this does not violate the standard Heisenberg relation,
which does not use inferred values. For a paradox achieved by condition
(\ref{eq:steer_m|c}), we can also conclude that pulse $c$ can \textquotedblleft{}steer\textquotedblright{}
the mechanical oscillator $m$ \cite{steering_wiseman,cava_steering2009}. 

\begin{figure}[t]
\begin{centering}
\includegraphics[width=0.9\columnwidth]{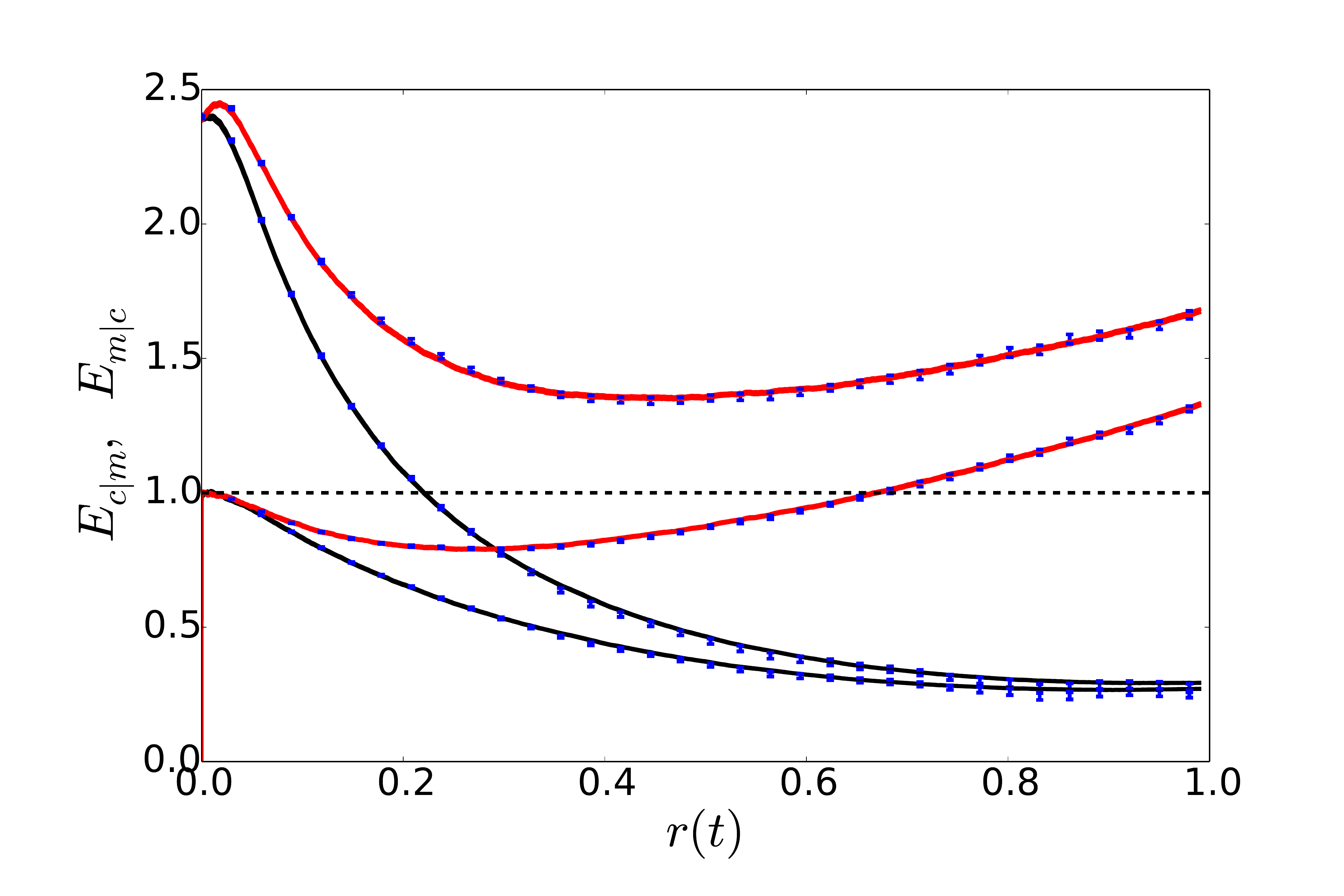}
\par\end{centering}

\caption{(Color online) Quantum EPR-steering versus squeezing parameter $r$,
for the same conditions used in Fig. 2. The two black curves are for
$T_{bath}=200mK$ ($\bar{n}_{b,th}=0.7$) and the two solid curves
are for a higher temperature heat bath with $T_{bath}=20K$ ($n_{b,th}=112$).
For lower bath temperatures, with a squeezing parameter $r\leq r_{0}$,
$E_{c|m}<1$ (lower black) means that only the mirror can steer the
optical system, while for $r>r_{0}$, $E_{m|c}<1$ (upper black curve)
means that both the mirror \emph{and} the pulse can steer the other
remote subsystem. For a higher temperature heat bath, only one-way
steering is possible. The presence of the thermal mechanical excitation
makes the steering of the oscillator by the pulse, a more difficult
challenge experimentally. \label{fig:one-way-epr}}
\end{figure}

A thermal barrier exists for this paradox. Figure \ref{fig:one-way-epr}
shows that the mechanical oscillator is steerable by the optical pulse
when $r>r_{0}$, where a minimum strength $r_{0}$ of the squeezing
parameter required, for a given $n_{b,0}$. A thermal barrier existing
means that a threshold level of pulse-oscillator interaction is required
for a given thermal occupation $n_{b,0}$ of the oscillator.

An EPR paradox can be shown the other way, by the criterion
\begin{align}
E_{c|m} & =4\Delta_{inf}\hat{X}_{c}^{out}\Delta_{inf}\hat{P}_{c}^{out}\nonumber \\
 & =4\Delta(\hat{X}_{c}^{out}+g_{x}\hat{X}_{m}^{out})\Delta(\hat{P}_{c}^{out}+g_{p}\hat{P}_{m}^{out})<1.\label{eq:steer_c|m}
\end{align}
Figure \ref{fig:one-way-epr} shown that this is possible for any
value of initial thermal noise $n_{b,0}$, and for any squeezing parameter
$r>0$. This means that there is no equivalent thermal barrier for
the optical pulse \textquotedblleft{}steered\textquotedblright{} by
the measurements made on the mechanical oscillator. This effect is
possible because the pulse is not thermally excited. We also find
that $E_{c|m}$ is less sensitive to mechanical decoherence. The reason
for this is that we can select optimal gain values $g$ to reduce
the effect of the initial thermal noises $n_{b,0}$ and $ $the heat
bath $n_{b,th}$.

This simulation of the full quantum dynamics not only validates previous
analytic results but includes an analysis of the full standard optomechanical
Hamiltonian, rather than an idealized parametric model. For a squeezing
parameter $r\leq r_{0}$, the only EPR paradox possible is that which
verifies the steering of the optical system by the measurements on
the mechanical one, i.e. ``one-way steering'' \cite{optoHe&Reid,one-way-NPho2012,one-way-Olsen2010,one-way-Wagner}
is predicted. For $r>r_{0}$, ``two-way steering'' becomes possible,
meaning that both mirror and pulse can steer the other subsystem.
Clearly, the asymmetry of the steering is due to the asymmetry of
the thermal effects on the two systems. The result suggests a fundamental
physical principle: \emph{changes to a massive system as a result
of an action of measurement at a distant site will be inhibited by
thermal noise}.  

\textbf{Summary} 

Optomechanics presents a challenge for exact quantum simulations.
It combines a range of occupation numbers and time-scales with non-equilibrium
and nonlinear open system quantum dynamics. These results demonstrate
that the exact positive-P representation approach is able to give
a first principles simulation of the standard model \cite{Pace-Collett-Walls}.
For the current parameters simulated here, the truncated Wigner approach
is also reliable, although this method needs to be verified by more
precise methods like the positive-P approach at large couplings. 

We find that probabilistic phase-space simulations of optomechanics
are practical and can be carried out with negligible sampling error.
An important consequence is that previously used approximations appear
unnecessary. For example, our simulations employed square pulses where
the adiabatic approximation fails completely - yet there is still
very strong entanglement and EPR steering in the simulations. Unlike
number-state based approaches, our method can be readily scaled up
to study multipartite systems with many oscillator modes \cite{Chan-optocrystal}.

\textbf{Acknowledgments.} - We acknowledge support from the Australian
Research Council via Discovery and DECRA grants.


\begin{thebibliography}{References}
\bibitem{cooling_Connell_2010_Nature}A. D. O'Connell et al., Nature
\textbf{464}, 697 (2010).

\bibitem{cooling_Chan_2011_Nature}J. Chan et al., Nature \textbf{478},
89 (2011). 

\bibitem{cooling_Gro_2009_NP}S. Groblacher et al., Nature Physics
\textbf{5}, 485-488 (2009).

\bibitem{cooling_Teufel}J. D. Teufel et al., Nature \textbf{475},
359 (2011); \textbf{471}, 204 (2011).

\bibitem{quantum_macro}A. Safavi-Naeini et al, Phys. Rev. Lett. \textbf{108},
033602 (2012).

\bibitem{quantum_macro2}N. Brahms et al., Phys. Rev. Lett. \textbf{108},
133601 (2012).

\bibitem{quantum_macro3}T. J. Kippenberg and K. J. Vahala, Science
\textbf{321}, 1172 (2008).

\bibitem{quantum_macro4}M. Aspelmeyer et al., J. Opt. Soc. Am. B
\textbf{27}, A189 (2010).

\bibitem{EPR1935}A. Einstein, B. Podolsky, and N. Rosen, Phys. Rev.
\textbf{47}, 777 (1935).

\bibitem{Reid1989}M. D. Reid, Phys. Rev. A \textbf{40}, 913 (1989).

\bibitem{Schrodinger}E. Schrödinger, Naturwissenschaften \textbf{23},
844 (1935). 

\bibitem{ent-opto-Giovannetti}V. Giovannetti, S. Mancini and P. Tombesi,
Europhys. Lett. \textbf{54}, 559 (2001).

\bibitem{ent-opto-Mancini}S. Mancini, V. Giovannetti, D. Vitali and
P. Tombesi, Phys. Rev. Lett. \textbf{88}, 120401 (2002).

\bibitem{Simon}W. Marshall, R. Simon, R. Penrose and D. Bouwmeester,
Phys. Rev. Lett., \textbf{91}, 130401 (2003).

\bibitem{Born}Max Born, \emph{Born-Einstein Letters, 1916-1955: Friendship,
Politics and Physics in Uncertain Times} (Palgrave MacMillan, 2005).

\bibitem{Schrodinger-steer}E. Schroedinger, Proc. Cambridge Philos.
Soc. \textbf{31}, 555 (1935); Proc. Cambridge Philos. Soc. \textbf{32},
446 (1936).

\bibitem{steering_wiseman}H. M. Wiseman, S. J. Jones, and A. C. Doherty,
Phys. Rev. Lett. \textbf{98}, 140402 (2007); S. J. Jones, H. M. Wiseman
and A. C. Doherty, Phys. Rev. A \textbf{76}, 052116 (2007). 

\bibitem{cava_steering2009}E. G. Cavalcanti et al., Phys. Rev. A
\textbf{80}, 032112 (2009).

\bibitem{ent-opto-Hofer}S. G. Hofer et al., Phys. Rev. A \textbf{84},
052327 (2011).

\bibitem{gen-p-rep} P. D. Drummond and C. W. Gardiner, J. Phys.\ A:
Math.\ Gen.\ \textbf{13}, 2353 (1980). We note that growth of sampling
errors in time limit this approach in undamped cases.

\bibitem{fast-coupling-Vanner2011}M. R. Vanner et al., Proc. Nat.
Ac. Sc. \textbf{108}, 16182 (2011).

\bibitem{fast-coupling-Pikovski-np2012}I. Pikovski et al., Nat. Phys.
\textbf{8}, 393 (2012).

\bibitem{DGCZ}L. M. Duan, G. Giedke, J. I. Cirac, and P. Zoller,
Phys. Rev. Lett. \textbf{84}, 2722 (2000).

\bibitem{Wigner}E. P. Wigner, Phys. Rev. \textbf{40} 749 (1932);
J.E. Moyal, Proc. Camb.e Phil. Soc. \textbf{45}, 99 (1949); R. Graham,
in \textit{Springer Tracts in Modern Physics: Quantum Statistics in
Optics and Solid-State Physics}, edited by G. Hohler, (Springer, New
York, 1973).

\bibitem{Wigner_Steel}M. J. Steel et al., Phys. Rev. A \textbf{58},
4824 (1998).

\bibitem{Wigner_Castin}A. Sinatra, C. Lobo, and Y. Castin, J. Phys.
B \textbf{35}, 3599 (2002).

\bibitem{Peterepl1993}P. D. Drummond and A. D. Hardman, Europhys.
Lett. \textbf{21}, 279 (1993).

\bibitem{Peter2002}S. Chaturvedi, K. Dechoum, and P. D. Drummond,
Phys. Rev. A \textbf{65}, 033805 (2002).

\bibitem{Dechoum2004}K. Dechoum et al., Phy. Rev. A \textbf{70},
053807 (2004).

\bibitem{Reid-Drumm}M.D. Reid, P.D. Drummond, Phys. Rev. A \textbf{40},
4493 (1989), P.D. Drummond, M.D. Reid, Phys. Rev. A \textbf{41}, 3930
(1990). 

\bibitem{Ludwig}Max Ludwig, Björn Kubala and Florian Marquardt, New
Journal of Physics \textbf{10} 095013 (2008); Max Ludwig, Amir H.
Safavi-Naeini, Oskar Painter, and Florian Marquardt, Phys. Rev. Lett.
\textbf{109}, 063601 (2012).

\bibitem{Kronwald}Andreas Kronwald, Max Ludwig, and Florian Marquardt,
Phys. Rev. A \textbf{87}, 013847 (2013); 

\bibitem{Qiu}Liu Qiu, Lin Gan, Wei Ding, and Zhi-Yuan Li, J. Opt.
Soc. Am. B \textbf{30}, 1683 (2013).

\bibitem{Carter}S. J. Carter, M. D. Reid, R. M. Shelby and P. D.
Drummond, Phys. Rev. Letters \textbf{58}, 1841 (1987). 

\bibitem{LLoyd}Seth Lloyd, Science \textbf{273}, 1073 (1996).

\bibitem{Opanchuk}Bogdan Opanchuk, Laura Rosales-Zárate, Margaret
D. Reid, Peter D. Drummond, arXiv:1301.6305.

\bibitem{Chan-optocrystal}Jasper Chan, Amir H. Safavi-Naeini, Jeff
T. Hill, Sean Meenehan, and Oskar Painter, Applied Phys. Letts. \textbf{101},
081115 (2012).

\bibitem{optoHe&Reid}Q. Y. He and M. D. Reid, Phys. Rev. A \textbf{88},
052121 (2013).

\bibitem{Hamil_Brag}V. B. Braginsky and A. Manukin, Sov. Phys. JETP
\textbf{25}, 653 (1967).

\bibitem{Hamil_pierre}A. Dorsel et al., Phys. Rev. Lett. \textbf{51},
1550 (1983).

\bibitem{Pace-Collett-Walls} A. F. Pace, M. J. Collett, D. F. Walls,
Phys. Rev. A \textbf{47}, 3173 (1993). This widely used model incorporates
the Markovian and small-displacement approximations, which are typically
very well satisfied.

\bibitem{inout_Yurke}B. Yurke, Phys. Rev. A \textbf{32}, 300 (1985).

\bibitem{inout-Gardiner-Collett}C. W. Gardiner and M. J. Collett,
Phys. Rev. A \textbf{31}, 3761 (1985); M. J. Collett and D. F. Walls,
ibid. \textbf{32}, 2887 (1985).

\bibitem{rmp_reid}M. D. Reid et al., Rev. Mod. Phys. \textbf{81},
1727 (2009).

\bibitem{one-way-NPho2012}V. Händchen et al., Nature Photonics \textbf{6},
596 (2012).

\bibitem{one-way-Olsen2010}S. Midgley, A. J. Ferris and M. K. Olsen,
Phys. Rev. A \textbf{81}, 022101 (2010). 

\bibitem{one-way-Wagner}K. Wagner et al., arXiv:1203.1980 {[}quant-ph{]}.\end{thebibliography}
\end{document}